\documentclass[twocolumn,showpacs,amsmath,nofootinbib]{revtex4-2}

\usepackage{float} 
\usepackage{amsmath}
\usepackage{amsfonts}
\usepackage{amssymb}
\usepackage{amsthm}
\usepackage{graphicx}
\usepackage[colorlinks,citecolor=blue]{hyperref}
\usepackage{color}
\usepackage{flafter}
\usepackage[normalem]{ulem}

\usepackage[toc,page]{appendix}
\usepackage{makecell}
\usepackage{slashed}
\usepackage{tikz-cd}
\usepackage{makecell}
\usepackage{subcaption}
\usepackage{graphicx}
\usepackage{xr}
\usepackage[font=small,labelfont=bf,
justification=justified,
format=plain]{caption}

\allowdisplaybreaks

\definecolor{OliveGreen}{rgb}{0,0.6,0}
\definecolor{Orange}{rgb}{1.00, 0.65, 0}
\definecolor{Grey}{rgb}{0.43, 0.5, 0.5}

\newcommand{\Fig}[1]{Fig.~\ref{#1}}
\newcommand{\Eq}[1]{Eq.(\ref{#1})}
\newcommand{\nn}{\nonumber\\}

\newcommand{\be}{\begin{eqnarray}}
\newcommand{\ee}{\end{eqnarray}}
\newcommand{\bpm}{\begin{pmatrix}}
\newcommand{\epm}{\end{pmatrix}}

\newcommand{\p}{\partial}

\newcommand{\ra}{\rightarrow}

\renewcommand{\a}{\alpha}

\newcommand{\e}{\epsilon}
\newcommand{\s}{\sigma}
\renewcommand{\t}{\tau}

\newcommand{\w}{{\omega}}

\newcommand{\comment}[1]{}

\begin{document}
\title{The electromagnetic duality and the 3+1D O(6) non-linear sigma model with a level-1 Wess-Zumino-Witten term}
\author{Yen-Ta Huang$^{1}$}
\author{Dung-Hai Lee$^{1,2}$}\email{Corresponding author: dunghai@berkeley.edu}

\affiliation{
$^1$ Materials Sciences Division, Lawrence Berkeley National Laboratory, Berkeley, CA 94720, USA.\\
$^2$ Department of Physics, University of California, Berkeley, CA 94720, USA.\\
}

\begin{abstract}
We show that in (3+1)-D space-time dimensions, the O(6) non-linear sigma model, with a level-1 Wess-Zumino-Witten term, exhibits the electromagnetic duality. If we name the six components of the sigma field as the Neel and valence-bond-solid (VBS) order parameters, the hedgehogs of the Neel and VBS order play the role of monopole and charge. The duality corresponds to the exchange of monopole and charge hence the exchange of the Neel and VBS order. The condensation of monopole can trigger a direct Neel $\leftrightarrow$ VBS phase transition. We conjecture that the critical point is self-dual, which is a generalization of the deconfined quantum critical point in (2+1)-D. In this theory, there exist two deconfined phases where the Neel/VBS hedgehogs are massive but cost finite energy. This leads to two fractionalized phases with particles carrying fractional spin or VBS quantum numbers and gapless gauge bosons. 
\end{abstract}
\maketitle

\newpage

\noindent{{\bf Introduction}} Non-linear sigma (NL$\s$) models describe the dynamics of Goldstone mode. Due to the interaction between the Goldston bosons, NL$\s$ models can develop a mass instead of being gapless \cite{Polyakov1975}. A well-known example occurs in 1+1 dimensions, where the $O(3)$ NL$\s$ model is always gapful. On a different front, topological terms in the NL$\s$ model can fundamentally modify the energy spectrum. For example, the $O(3)$ NL$\s$ model with $\theta$ term is gapful for $\theta=2n \pi$ while it is gapless for $\theta=(2n+1 )\pi$ for $n\in \mathbb{Z}$ \cite{Haldane1983}. In addition to the topological $\theta$ term, other topological terms, such as the Wess-Zumino-Witten \cite{Witten1984} and the Hopf terms \cite{Polyakov1988}, are also known to change the Goldstone mode spectrum.

In Ref.\cite{Senthil2004}, Senthil {\it et al.} proposed that despite the mismatch in symmetries, there can exist a continuous phase transition, which is dubbed ``deconfined quantum critical point'', between the Neel and the valence-bond-solid (VBS) states in quantum magnets. Shortly after, it is proposed that the O(5) (3 Neel+2 VBS) NL$\s$ model with a level-1 Wess-Zumino-Witten (WZW) term describes such a phase transition \cite{Tanaka2005}. Recently, we point out that the same model can give rise to spin liquid and spin rotation symmetry breaking phases between the Neel and VBS order \cite{Huang2022}. Moreover, an unexpected connection between the NL$\s$ model approach, and the Schwinger boson approach \cite{Sachdev1991} was revealed. This connection is interesting because, in the Schwinger boson approach, particles that fractionalize spins were built-in from the beginning. However, aside from the WZW term, the NL$\s$ model in Ref.\cite{Huang2022} only involves conventional order parameters. This connection between the WZW term and fractionalized particles is very intriguing, and it is natural to ask whether it can occur for higher dimensions.

\noindent{{\bf The $O(6)$ NL$\s$ model with a level-1 Wess-Zumino-Witten (WZW) term in 3+1 space-time dimensions}} The action of this model read
\be &&S={1\over 2g}\int\limits_{\mathcal{M}} d^4 x \, \left( \partial_\mu \Omega_i \right)^2-W_{\rm WZW}[\tilde{\hat{\Omega}}]\label{o61}\ee
\be W_{\rm WZW}[\tilde{\hat{\Omega}}]&&=\frac{2\pi i }{120 \pi^3} \int\limits_{\mathcal{B}}~ \epsilon^{ijklmn}~\tilde{\Omega}_i \, d\tilde{\Omega}_j d\tilde{\Omega}_k\nn&& d\tilde{\Omega}_l d\tilde{\Omega}_md\tilde{\Omega}_n.\nn\label{o6}\ee
For easy reference, we shall name the first three components of $\hat{\Omega}$ the Neel order parameters, and the last three components the valence bond solid (VBS) order parameters, i.e., $\hat{\Omega}=(n_1,n_2,n_3,v_1,v_2,v_3)$. These names are motivated by the fact that \Eq{o6} governs the low-energy dynamics of a three-dimensional Mott insulator with competing Neel and valence-bond solid (VBS) order \cite{Huang2021} (see supplemental material section I
). However, for the majority of this paper, these are just names we use to refer to the components of the $\hat{\Omega}$. 

In \Eq{o6} the space-time manifold $\mathcal{M}=S^4$ is spanned by $\t,x,y,z$ and $\mathcal{B}$ is spanned by $\t,x,y,z$ and $u\in [0,1]$ such that $\p\mathcal{B}=\mathcal{M}.$ $\tilde{\hat{\Omega}}(\tau,x,y,u)$ represents a one-parameter-family extension of the $\hat{\Omega}(\t,x,y,z)$ such that at $u=0$, the configuration is trivial, say, $\tilde{\hat{\Omega}}(\tau,x,y,z,0)=(0,0,0,0,1)$, and at $u=1$ the $\hat{\Omega}(\t,x,y,z,1)=\hat{\Omega}(\t,x,y,z)$. It can be shown that $\exp\left(-W_{WZW}\right)$ is independent of $\tilde{\hat{\Omega}}$ for $u<1$ as long as the coefficient in front of the WZW term is an integer multiple of $\frac{2\pi i }{120 \pi^3}$. 

\noindent{{\bf The VBS hedgehog}}\label{vhedgehog}
Consider the following $\tilde{\hat{\Omega}}$ configuration corresponding to a hedgehog in $\left(v_1,v_2,v_3\right)$ 
\be
&&(\tilde{\Omega}_4,\tilde{\Omega}_5,\tilde{\Omega}_6)=\sin f(r) (\sin\theta\cos\phi,\sin\theta\sin\phi,\cos\theta)\nn
&&(\tilde{\Omega}_1,\tilde{\Omega}_2,\tilde{\Omega}_3)=\cos f(r) (n_1(u,\t),n_2(u,\t),n_3(u,\t)),
\label{hedgehog}
\ee
where $(r,\theta,\phi)$ are the coordinates of the three dimensional space and $\t$ is the Euclidean time. In \Eq{hedgehog}, $(n_1,n_2,n_3)$ is a unit vector and $f(r)$ is a smooth function satisfying $f(r)=\pi/2$ for $r>r_c$ (hedgehog core size) and $f(r)=0$ at $r=0$. It is trivial to show that $\sum_{i=1}^6\tilde{\Omega}_i^2=1.$
Plug \Eq{hedgehog} into \Eq{o6}, it is straightforward to show that 
\be
W_{\rm WZW}\ra W^{\text{VBS hh}}_{\rm WZW}={2\pi i\over 8\pi}\int ~\e^{ijk}n_i dn_j dn_k.\label{spinB}\ee
\Eq{spinB} is the Berry's phase in the coherent state path integral of a spin 1/2 in 0+1 dimension. Therefore the hedgehog of the VBS order fractionalizes the spins! In the supplemental material section II
, we present the microscopic theory of the hedgehog core.

If we define a two-component complex boson field 
\be z^T=(z_1,z_2)~~{\rm where~~} |z_1|^2+|z_2|^2=1,\label{z}\ee so that \be\hat{n}=z^\dagger\vec{\s}z,\label{redg}\ee it is simple to show that\cite{Huang2022} 
\be
W_{\rm WZW}^{\text{VBS hh}}=i\times \int d\tau\left({1\over i}z^\dagger\p_\t z\right).\label{sb}\ee
More generally, for a dynamic hedgehog where the world line forms a closed loop parametrized by $\zeta$, \Eq{sb} becomes
\be
W_{\rm WZW}^{\text{VBS hh}}\ra i\times \int d\zeta\left({1\over i}z^\dagger\p_\zeta z\right).\label{sb1}\ee
This implies the hedgehog current $J_\mu$ couples to a gauge field, namely, 
$$i \int d^4 x J^\mu a_\mu~~{\rm where}~~ a_\mu={1\over i}z^\dagger\p_\mu z. $$

\noindent{{\bf The statistics of VBS hedgehogs}}
We can determine the statistics of the VBS hedgehogs by computing the Berry phase associated with the hedgehog exchange. Because our space-time is $S^4$, the exchange process must be embedded in (i) vacuum creation of two pairs of hedgehog-anti-hedgehog, (ii) exchanging the two hedgehogs, and (iii) annihilating the hedgehogs with the anti-hedgehogs. The Berry phase contains two contributions: 1) that due to exchange of the hedgehogs and 2) that associated with the spin 1/2 in the cores of hedgehog and anti-hedgehog. To isolate the Berry phase due to the exchange, we lock the core spins in, say, the positive $n_3$-direction.

In the following, we present an argument suggesting that under the space-time configuration discussed above, the Berry phase due to hedgehog exchange is zero. The argument involves two steps: (i) using the result in Ref.\cite{Abanov2000}, one can show that when any {\it one} of the six components of $\hat{\Omega}$ is zero, the WZW term reduces to the topological $\theta$-term (with $\theta=\pi$) for the remaining five components. This topological term is non-zero only when the wrapping number associated with the mapping from the space-time to the order parameter manifold formed by the non-zero components, in this case, $S^4$, is non-zero. (ii) We note that in the present situation only four components of $\hat{\Omega}$ are non-zero. By counting the dimension of the space-time image, we conclude that any such $\hat{\Omega}$ cannot produce a non-zero wrapping number, hence the Berry phase vanishes. This argument suggests that the VBS hedgehogs are bosons. A similar argument can be used to deduce that the Neel hedgehog is bosons as well.

\noindent{ {\bf The CP$^1$ theory of Neel order}}\label{ne}
The NL$\s$ model action for the Neel order parameter reads
\be
S_{\rm AF}={1\over 2g_{\rm AF}}\int d^4 x (\p_\mu\hat{n})^2.\label{nnls}\ee It is well known that the above action can be rewritten as\cite{Witten1979}
\be
S_{\rm AF}={1\over 2g_{\rm AF}}\int d^4 x |(\p_\mu-ia_\mu)z|^2.\label{cp1}\ee where $a_\mu$ is a compact dynamic gauge field and $z$ is the two-component complex boson field satisfying \Eq{z} and \Eq{redg}. Because the action in \Eq{cp1} is at most quadratic in $a_\mu$ the saddle point is exact, namely, \be {\delta S_{\rm AF}\over \delta a_\mu}=0\Rightarrow a_\mu={1\over i}z^\dagger\p_\mu z.\label{eoma}\ee In \Eq{cp1}, the condensation of the $z$ boson Higgs the gauge field $a_\mu$, hence mods out the redundant local phase gauge of freedom $z\ra z^{i\varphi}z$ in \Eq{redg}. The resulting phase exhibits the Neel long-range order. Combining \Eq{sb1} and the
above discussion suggests that the field theory for the VBS hedgehog is given by
\be
S_{\rm VBS~hh}=\int d^4 x \left\{{1\over 2g_1}|(\p_\mu-ia_\mu)z|^2+{1\over 2g_2}(\e^{\mu\nu\rho}\p_\nu a_\rho)^2\right\}.\nn\label{VBShedgehog}\ee
\noindent Here $z$ represents the boson field for the VBS hedgehog. Once again, we see that when supplemented with the WZW term, the conventional Ginzburg-Landau theory is equivalent to a theory with fractionalized particles and gauge field. 

\noindent{{\bf The Neel hedgehog}}\label{nhedgehog}
A parallel discussion with ``VBS'' and ``Neel'' switched implies that the Neel hedgehog carries the Berry phase of the VBS order parameter. Specifically after the switch \Eq{spinB} becomes 
\be
W^{\text{Neel~hh}}_{\rm WZW}= {2\pi i\over 8\pi}\int~\e^{ijk}v_i dv_j dv_k.\label{vbsb}\ee
If we define a two-component complex boson field 
\be w^T=(w_1,w_2)~{\rm where~} |w_1|^2+|w_2|^2=1,~{\rm and~}\hat{v}=w^\dagger\vec{\s}w,\nn\label{w}\ee it is simple to show that 
\be
W_{\rm WZW}^{\text{Neel~hh}}= i\times \int d\t\left({1\over i}w^\dagger\p_\t w\right).\label{vb}\ee
This implies the VBS hedgehog current $K_\mu$ couples to a gauge field, namely, 
\be i \int d^4 x K^\mu b_\mu~~{\rm where}~~ b_\mu={1\over i}w^\dagger\p_\mu w. \label{nhhc}\ee\\

\noindent{ {\bf The CP$^1$ theory of VBS order}}\label{vb}
The NL$\s$ model action for the VBS order parameter read
\be
S_{\rm VBS}={1\over 2g_{\rm VBS}}\int d^4 x (\p_\mu\hat{v})^2.\label{vnls}\ee Similar to \Eq{cp1} we can rewrite \Eq{vnls} as
\be
S_{\rm VBS}={1\over 2g_{\rm VBS}}\int d^4 x |(\p_\mu-ib_\mu)w|^2.\label{cp2}\ee where $b_\mu$ is a compact dynamic gauge field and $w$ is the two-component complex boson field satisfying \Eq{w}. In this case \be {\delta S_{\rm VBS}\over \delta b_\mu}=0\Rightarrow b_\mu={1\over i}w^\dagger\p_\mu w.\label{eoma2}\ee Analogous to the discussion after \Eq{cp1}, the condensation of the $w$ bosons induces the VBS long-range order. The above discussion suggests that the field theory for the Neel hedgehog is given by
\be
S_{\rm Neel~hh}&&=\int d^4 x\left\{ {1\over 2g_3}|(\p_\mu-ib_\mu)w|^2+{1\over 2g_4}(\e^{\mu\nu\rho}\p_\nu b_\rho)^2\right\}.\nn\label{Neelhedgehog}\ee
\noindent Here $w$ represents the boson field for the Neel hedgehog.\footnote{When \Eq{Neelhedgehog} is used to describe the acutal VBS order of a Mott insulator, there is an additional term $$S_{\rm aniso}[w^\dagger\vec{\s}w]$$ which should be added to \Eq{Neelhedgehog}. This term describes the anisotropy imposed by the lattice of the Mott insulator. However, for our purpose the term Neel and VBS are just names we assign to the components of $\hat{\Omega}$, therefore the above anisotropy term is absent.}

\noindent{{\bf The monopoles}}\label{mono}
One can ask what do the Neel and VBS hedgehog correspond to in terms of the fields in \Eq{VBShedgehog} and \Eq{Neelhedgehog}. Consider a static Neel hedgehog at the spatial origin, namely,
$$\hat{n}(r,\theta,\phi)=(\sin\theta\cos\phi,\sin\theta\sin\phi,\cos\theta)$$ where $r,\theta,\phi$ are the spatial spherical coordinates. Through $\hat{n}=z^\dagger\vec{\s}z$, the corresponding $z$ is\footnote{Here due to the singular configuration of $\hat{n}$ we need to take two patches to avoid the singularity around $\theta=0$ and $\theta=\pi$. Specifically the northern hemisphere patch is $$z^N(r,\theta,\phi)=\begin{pmatrix} 
\cos{\theta\over 2} \\
\sin{\theta\over 2}e^{i\phi} \\
\end{pmatrix},$$ and the southern hemisphere patch is $$z^S(r,\theta, \phi)=\begin{pmatrix} 
\cos{\theta\over 2}e^{-i\phi} \\
\sin{\theta\over 2} \\
\end{pmatrix}.$$ These two patches are related by a phase transformation $z^N=e^{i\phi} z^S$. In terms of $a_j={1\over i}z^\dagger\p_j z$ the phase transformation is a gauge transformation. However in the following we shall compute the field strength $f_{ij}=\p_ia_j-\p_ja_i$ which is independent of the gauge. Therefore we can use either gauge to do the calculation. }
$$z(r,\theta,\phi)=\begin{pmatrix} 
\cos{\theta\over 2} \\
\sin{\theta\over 2}e^{i\phi} \\
\end{pmatrix}.$$ 
\Eq{eoma} implies the corresponding gauge field (one-form) is $a={1\over i}z^\dagger d z$ whose field strength (two-form) is
\be f=da={1\over 2}\sin\theta ~d\theta d\phi.\label{2f}\ee If we integrate \Eq{2f} over any closed surface enclosing the hedgehog the result is $2\pi$. Hence the Neel hedgehog is a monopole in $a_\mu$. The same calculation goes through if we change Neel to VBS and $a_\mu$ to $b_\mu$, i.e., a VBS hedgehog is a monopole in $b_\mu$.

\noindent{{\bf The electromagnetic duality}}\label{em}
These discussions suggest the monopole in $b_\mu$ is the charge of $a_\mu$ and the monopole in $a_\mu$ is the charge of $b_\mu$. We summarize such relation in Table \ref{emd}.
\begin{table}
\begin{tabular}{ |c|c|c| }
\hline
~~ & Neel& VBS \\
\hline
Gauge field &$a_\mu$ &$b_\mu$\\
\hline
VBS hedgehog &charge&monopole\\
\hline
Neel hedgehog &monopole &charge\\
\hline
Hedgehog field &$\begin{pmatrix} 
z_1 \\
z_2 \\
\end{pmatrix}$ &$\begin{pmatrix} 
w_1 \\
w_2 \\
\end{pmatrix}$\\
\hline
\end{tabular}
\caption{The electromagnetic duality of the $O(6)$ NL$\s$ model.}\label{emd}
\end{table}
In this duality $a_\mu\leftrightarrow b_\mu$, $z\leftrightarrow w$ and Neel order $\leftrightarrow$ VBS order. Here $\leftrightarrow$ means ``is dual to''.

\noindent{{\bf The Neel to VBS phase transition}}
In the above duality, the condensation of, e.g., the $b_\mu$ monopole will confine the $b_\mu$ charge. According to Table \ref{emd} the $b_\mu$ charge is the $a_\mu$ monopole. Putting it differently, since the $b_\mu$ monopole is the $a_\mu$ charge, condensation of it will Higgs $a_\mu$ hence confine the $a_\mu$ monopole. Of course, the statement with $a_\mu$ and $b_{\mu}$ exchanged is true. If there is a direct continuous phase transition between the Neel (condensation of $b_\mu$ monopole or $a_\mu$ charge ) and VBS (condensation of $b_\mu$ charge or $a_\mu$ monopole) order, which is equivalent to a direct transition from the confinement to the Higgs phase in \Eq{VBShedgehog} and \Eq{Neelhedgehog}. At the critical point, neither the VBS nor the Neel hedgehog can condense but they can be both gapless. We conjecture that this happens at the self-dual point of the electromagnetic duality discussed in Table \ref{emd}. The possibility of such a direct, continuous phase transition is supported by the Monte-Carlo simulation of the lattice version of \Eq{VBShedgehog} \cite{Sawamura2008}.
\begin{figure}
\includegraphics[scale=0.25]{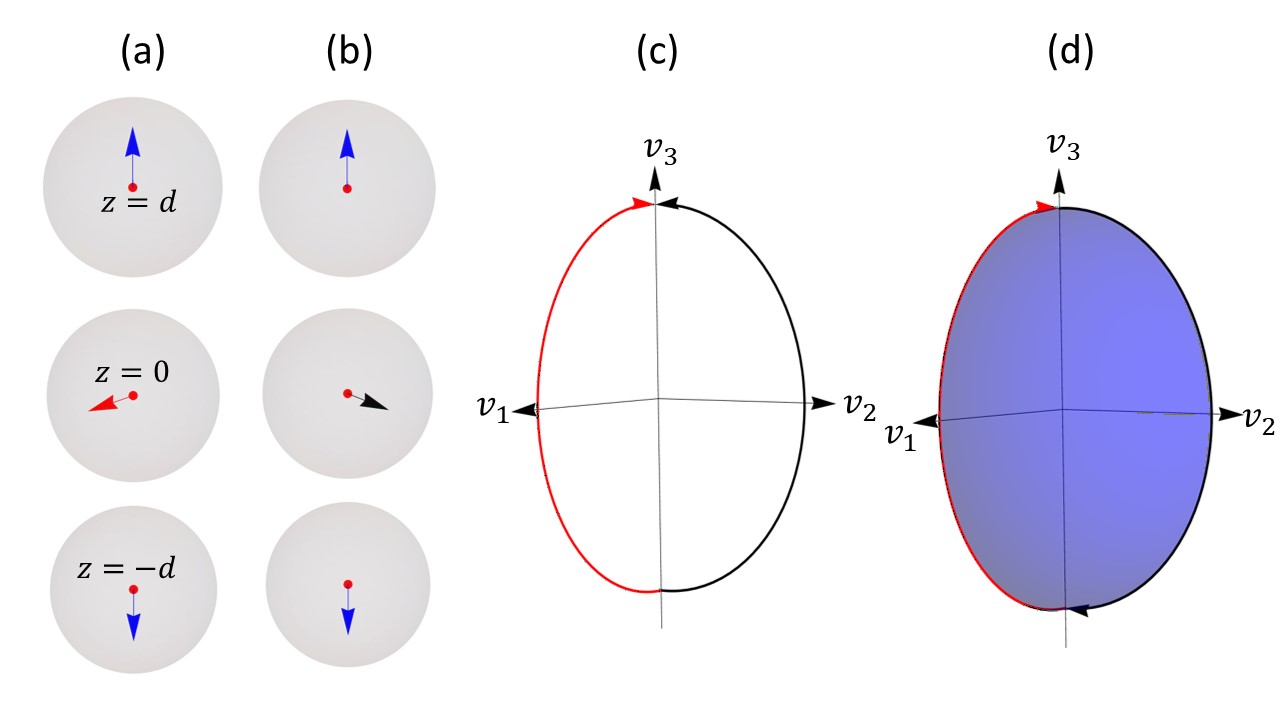}
\caption{(a,b) A $v_3$ (blue arrow) domain wall situated at $z=0$. The gray spheres schematically represent the Neel hedgehog. The hedgehog sitting in the domain wall ($z=0$) has a red (a) and back (b) arrow in the core. This indicates the direction of the VBS order parameter. (c) The red and black curves represent the trajectories traversed by the core VBS order parameter in part (a) and part (b). (d) The Berry phase difference between (a) and (b) is one-half of the solid angle sustained by the blue surface. } 
\label{dw}
\end{figure}

\noindent{{\bf Spin liquids}}
In Ref.\cite{Sawamura2008}, it is also shown that in the plane spanned by $g_1$-$g_2$ in \Eq{VBShedgehog}, there exists a deconfined Coulomb phase where $a_\mu$ gauge bosons are gapless.
In this phase the $a_\mu$ charge or $b_\mu$ monopole is massive, but finite energy, excitations. Since the $a_\mu$ charge is the spinon, this phase is a spin liquid. A similar argument applies to \Eq{Neelhedgehog}. In the deconfined phase, the $b_\mu$ gauge bosons are gapless, and the $b_\mu$ charge is a finite energy excitation. Since the $b_\mu$ charge fractionalizes the VBS order parameter, this phase is a ``VBS'' liquid.
The spin liquid phase discussed above is analogous to that found in Ref.\cite{Motrunich2004} for O(3) NL$\s$ model in (2+1)-D after the hedgehog suppression. In the phase diagram, these two fractionalized phases should sit symmetrically about the self-dual point.

The  fractionalized phases discussed above are made possible by the fact that in 3+1 dimensions, the compact U(1) gauge theory has a deconfinement phase. This is very different from 2+1 dimensions, where compact U(1) gauge theory always confines \cite{Polyakov1977}. In the latter case, deconfinement requires the condensation of an additional Higgs field carrying, e.g., two units of the $a_\mu$. This Higgs the gauge group of $a_\mu$ to $\mathbb{Z}_2$ which has a deconfined phase. \\
\noindent{{\bf Dimension reduction and the relation to O(5) NL$\s$ model with WZW term in 2+1D}}\label{dre}
In general, given an $O(n)$ NL$\s$ model with WZW term in $D$ space-time dimensions, we can obtain an $O(n-1)$ NL$\s$ model with WZW term in $D-1$ space-time dimensions by the following dimension reduction. The idea is to create a domain wall in the, say, last component of $\hat{\Omega}$. Specifically let us consider the following $\tilde{\hat{\Omega}}$ configuration where 
\be
&&\tilde{\Omega}_6=\sin f(z)\nn
&&\tilde{\Omega}_j=\cos f(z)\tilde{\omega}_j(\t,x,y,u)~~j=1,...,5.
\label{wll}\ee
Here $f(z)$ is a smooth function satisfying $f(z)={\pi\over 2}~{\rm sign}(z)$ for $|z|>d$ (the domain wall thickness) and $f(z)=0$ at $z=0$. Plug \Eq{wll} into \Eq{o6} it is straightforward to show that $W_{\rm WZW}\ra W^{\rm dw}_{\rm WZW}$ where
\be
W^{\rm dw}_{\rm WZW}={2\pi i\over 64\pi^2}\int \e^{abcde}\tilde{\w}_a d\tilde{\w}_b d\tilde{\w}_c d\tilde{\w}_d d\tilde{\w}_e\nn
\label{wll1}\ee
\Eq{wll1} is precisely the WZW term of $O(5)$ NL$\s$ model in (2+1)-D. Adopting the name that $(n_1,n_2,n_3,v_1,v_2,v_3)$ are the components of the $O(6)$ NL$\s$ model, and $(n_1,n_2,n_3,v_1,v_2)$ are the components of the $O(5)$ NL$\s$ model, we next show that the space-time Neel hedgehog in the $O(5)$ NL$\s$ model carries a Berry's phase that depends on the direction of the VBS order parameter $(v_1,v_2)$ in the core. 

Consider the space-time configuration of the $O(6)$ NL$\s$ model shown in \Fig{dw}(a),(b). Here a $2+1$ dimensional domain wall in $v_3$ spanned by $\t$-$x$-$y$ with width $d$ is centered at $z=0$. A Neel hedgehog sits in the domain wall region with the core VBS order parameter points in different directions at $z=0$ as shown in \Fig{dw}(a) and (b). According to \Eq{nhhc} the hedgehog couples to the Berry connection ${1\over i} w^\dagger\p_z w$ (see section III 
of the supplemental material for more details). In the order parameter space spanned by $v_1$-$v_2$-$v_3$, the core VBS order parameter in \Fig{dw}(a) and (b) traces out the red and black trajectories, respectively (see \Fig{dw}(c)). The Berry phase difference is one-half of the soliton angle sustained by the blue surface shown in \Fig{dw}(d). This is precisely the hedgehog Berry's phase discussed by Haldane \cite{Haldane1988}. In his case the x-y plane is a square lattice favoring 4 different VBS angles $\varphi$ ($v_1+iv_2=e^{i\varphi}$), namely, $\varphi=\a,\a+\pi/2,\a+\pi,\a+3\pi/2.$ ($\a=0$ corresponds to the columnar VBS phase and $\a=\pi/4$ the plaquette VBS phase). This gives rise to four relative hedgehog Berry's phase, namely, $0,\pi/2,\pi,3\pi/2$, between the hedgehog with different VBS angles in the core, which causes destructive interference when hedgehog proliferate. The exception is hedgehogs with topological charge 4, which have the same Berry's phase regardless of the core VBS order parameter. When such degree-4 hedgehogs proliferate and condense, it results in the four-fold degenerate VBS order. \\
{\bf{Conclusion:}} In this paper we demonstrate that the O(6) NL$\s$ model in (3+1)-D exhibits the electromagnetic duality. Novel phase transition and fractionalized phases associated with this duality are discussed.

\bibliographystyle{ieeetr}
\bibliography{bibs_O6}

\begin{thebibliography}{10}

\bibitem{Polyakov1975}
A.~Polyakov, ``Interaction of goldstone particles in 2 dimensions -
  applications to ferromagnets and massive yang-mills fields,'' {\em Phys.
  Lett. B}, vol.~59, no.~1, pp.~79--81, 1975.

\bibitem{Haldane1983}
F.~Haldane, ``Continuum dynamics of the 1-{D} {H}eisenberg anti-ferromagnet -
  identification with the {O}(3) non-linear sigma-model,'' {\em Phys. Lett. A},
  vol.~93, no.~9, pp.~464--468, 1983.

\bibitem{Witten1984}
E.~Witten, ``Non-abelian bosonization in 2 dimensions,'' {\em Commun. Math.
  Phys.}, vol.~92, no.~4, pp.~455--472, 1984.

\bibitem{Polyakov1988}
A.~Polyakov, ``{F}ermi-{B}ose transmutations induced by gauge-fields,'' {\em
  Mod. Phys. Lett. A}, vol.~3, no.~3, pp.~325--328, 1988.

\bibitem{Senthil2004}
T.~Senthil, A.~Vishwanath, A.~Balents, S.~Sachdev, and M.~Fisher, ``Deconfined
  quantum critical points,'' {\em Science}, vol.~303, no.~5663, pp.~1490--1494,
  2004.

\bibitem{Tanaka2005}
A.~Tanaka and X.~Hu, ``Many-body spin berry phases emerging from the $\pi$-flux
  state: Competition between antiferromagnetism and the valence-bond-solid
  state,'' {\em Phys. Rev. Lett.}, vol.~95, no.~3, 2005.

\bibitem{Huang2022}
Y.-T. Huang and D.-H. Lee, ``{Competing orders, the {W}ess-{Z}umino-Witten
  term, and spin liquids},'' {\em {arXiv:2204.12485v1}}, 2022.

\bibitem{Sachdev1991}
S.~Sachdev and N.~Read, ``Large {N} expansion for frustrated and doped quantum
  antiferromagnets,'' {\em Int. J. Mod. Phys. B}, vol.~05, no.~01n02,
  pp.~219--249, 1991.

\bibitem{Huang2021}
Y.-T. Huang and D.-H. Lee, ``Non-abelian bosonization in two and three spatial
  dimensions and applications,'' {\em Nucl. Phys. B}, vol.~972, p.~115565,
  2021.

\bibitem{Abanov2000}
A.~Abanov and P.~Wiegmann, ``Theta-terms in nonlinear sigma-models,'' {\em
  Nucl. Phys. B}, vol.~570, no.~3, pp.~685--698, 2000.

\bibitem{Witten1979}
E.~Witten, ``Instantons, the quark-model, and the 1/{N} expansion,'' {\em Nucl.
  Phys. B}, vol.~149, no.~2, pp.~285--320, 1979.

\bibitem{Sawamura2008}
K.~Sawamura, T.~Hiramatsu, K.~Ozaki, I.~Ichinose, and T.~Matsui,
  ``Four-dimensional {$CP^1$+$U(1)$} lattice gauge theory for three-dimensional
  antiferromagnets: Phase structure, gauge bosons, and spin liquid,'' {\em
  Phys. Rev. B}, vol.~77, no.~22, 2008.

\bibitem{Motrunich2004}
O.~Motrunich and A.~Vishwanath, ``Emergent photons and transitions in the
  {O}(3) sigma model with hedgehog suppression,'' {\em Phys. Rev. B}, vol.~70,
  no.~7, 2004.

\bibitem{Polyakov1977}
A.~Polyakov, ``Quark confinement and topology of gauge theories,'' {\em Nucl.
  Phys. B}, vol.~120, no.~3, pp.~429--458, 1977.

\bibitem{Haldane1988}
F.~Haldane, ``O(3) nonlinear sigma-model and the topological distinction
  between integer-spin and half-integer-spin antiferromagnets in 2
  dimensions,'' {\em Phys. Rev. Lett.}, vol.~61, no.~8, pp.~1029--1032, 1988.

\end{thebibliography}


\begin{thebibliography}{1}

\bibitem{Affleck1988}
I.~Affleck and J.~Marston, ``Large-$n$ limit of the {H}eisenberg-{H}ubbard
  model: Implications for high-{$T_c$} superconductors,'' {\em Phys. Rev. B},
  vol.~37, p.~3774, 1988.

\bibitem{Huang2021}
Y.-T. Huang and D.-H. Lee, ``Non-abelian bosonization in two and three spatial
  dimensions and applications,'' {\em Nucl. Phys. B}, vol.~972, p.~115565,
  2021.

\bibitem{Huang2022}
Y.-T. Huang and D.-H. Lee, ``{Competing orders, the {W}ess-{Z}umino-Witten
  term, and spin liquids},'' {\em {arXiv:2204.12485v1}}, 2022.

\end{thebibliography}

\end{document}


\title{Supplemental materials: 3+1D O(6) non-linear sigma model with a level-1 Wess-Zumino-Witten term, and the electromagnetic duality} 
\author{Yen-Ta Huang$^{1}$}
\author{Dung-Hai Lee$^{1,2}$}\email{Corresponding author: dunghai@berkeley.edu}

\affiliation{$^1$ Department of Physics, University of California, Berkeley, CA 94720, USA.\\
$^2$ Materials Sciences Division, Lawrence Berkeley National Laboratory, Berkeley, CA 94720, USA.
}
\maketitle
\section{The 3d generalization of the Affleck-Marston $\pi$ flux phase and the O(6) NL$\s$ model with the level-1 WZW term}\label{pif}
\begin{figure}
\includegraphics[scale=0.35]{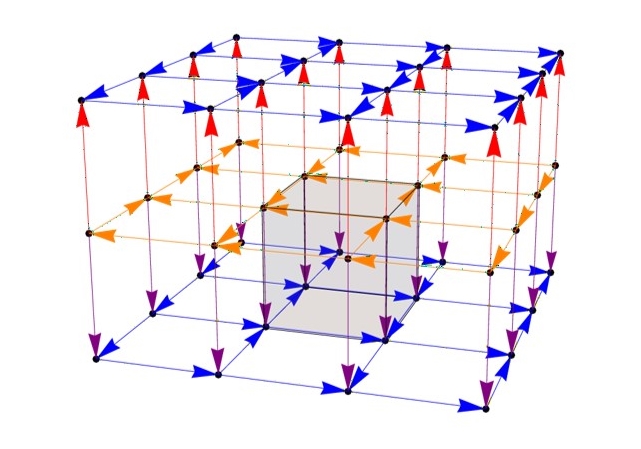}
\caption{The cubic $\pi$ flux model. The arrows point in the direction where the hopping matrix element is $+i$. The light gray region marks the unit cell.} 
\label{cubepi}
\end{figure}

The following 3D fermion lattice model, whose dispersion exhibits 3D Dirac nodes, is the generalization of the Affleck-Marston $\pi$ flux phase\cite{Affleck1988}. The hopping terms of this model are shown in \Fig{cubepi}, where the arrows point in the direction where the hopping matrix element is $+i$, and the light gray region marks the unit cell. Close inspection of \Fig{cubepi} reveals that there is a $\pi$ flux in every plaquette. In momentum space, such one-particle hopping Hamiltonian is given by
\be
h(\v k)=\sin k_1 ZIX +\sin k_2 ZXZ+\sin k_3 XII-(1-\cos k_1)ZIY-(1-\cos k_2)ZYZ-(1-\cos k_3)YII. \label{h}
\ee 
In \Eq{h} $I,X,Y,Z$ stands for Pauli matrices $\s_0,\s_x,\s_y,\s_z$ and the tensor product of Pauli matrices in \Eq{h} acts on the 8 sites in the unit cell shown in \Fig{cubepi}, and we have so far omitted the spin degrees of freedom. Including the spin, the Majorana representation of the lattice Hamiltonian is given by 
\be 
H=\sum_{\v k}\psi^T(-\v k) \left[II\otimes h(\v k)\right]\psi(\v k).\ee Here the first Pauli matrix acts on the Majorana index and the second on the spin index. 
The low energy Hamiltonian near the Dirac node is given by
\be 
H=\int d^3 x~\psi^T(\v x)\left[\sum_{i=1}^3 -i\G_i\p_i\right]\psi(\v x)\ee where
\be 
\G_1=IIZIX,~\G_2=IIZXZ,~\G_3=IIXII.\ee
In the following we focus on the six charge $SU(2)$ invariant \cite{Huang2021}\footnote{The charge $SU(2)$ generators are given by
$(XYIII, ZYIII,-YIIII)$.} mass terms given by
\be
\Omega_i=(n_1,n_2,n_3,v_1,v_2,v_3)\leftrightarrow M_i=(YXZZZ, IYZZZ,YZZZZ,IIZIY,IIZYZ,IIYII).\ee Here the first three are the VBS, and the second three are the Neel order parameters.
Straightforward integration of the gapped fermions\cite{Huang2021} yield the O(6) NL$\s$ model with a level-1 WZW term.
\be &&S={1\over 2g}\int\limits_{\mathcal{M}} d^4 x \, \left( \partial_\mu \Omega_i \right)^2-W_{\rm WZW}[\tilde{\hat{\Omega}}]\nn
&&W_{\rm WZW}[\tilde{\hat{\Omega}}]=\frac{2\pi i }{120 \pi^3} \int\limits_{\mathcal{B}}~ \epsilon^{ijklmn}~\tilde{\Omega}_i \, d\tilde{\Omega}_j d\tilde{\Omega}_k d\tilde{\Omega}_l d\tilde{\Omega}_md\tilde{\Omega}_n.\nn\label{o6}\ee
\\
\section{The fermion zero mode in the VBS hedgehog core}\label{hhcore}
The fermion zero modes in the VBS hedgehog core satisfy the equation
\be 
&&h_0 \cdot \phi(x,y,z)=0~~{\rm where}\nn&&h_0 =\sum_{i=1}^3\left[ -i\G_i\p_i+V_i(x,y,z)M_{3+i}\right].\ee 
Near the center of the hedgehog (which we choose to be the origin), $V_i\sim x_i$. Note that $h_0$ anticommutes with the matrix
$$D=-i~\G_1\G_2\G_3M_4M_5M_6=IIZZZ.$$ Hence in the basis that diagonalizes $D$, $h_0$ appears purely off-diagonal, i.e.,
\be h_0\ra \begin{pmatrix}0&Q\cr Q^\dagger&0\end{pmatrix} .\ee The solution of the zero mode $\phi=\begin{pmatrix}u\cr v\end{pmatrix}$ satisfies 
$$Qv=0~~{\rm or}~~Q^\dagger u=0$$ whichever normalizable. Note that $D=IIZZZ=\pm 1$ labels the A/B sub-lattice of the cubic lattice. 
A straightforward generalization of the analysis in Ref.\cite{Huang2022} shows that the hedgehog with $\pm 1$ topological charge has the fermion zero modes residing on the A or B sublattices. Each of these zero modes is spin degenerate. In the hedgehog core, turning on the Neel order parameter gaps out the zero-mode but yields spin 1/2 in the core. 

To analyze the hedgehog of the Neel order parameter, all we need to do is to perform the transformation using the following unitary matrix
\be
U=\exp\left({\pi\over 2}M_1M_4\right) \exp\left({\pi\over 2}M_2M_5\right) \exp\left({\pi\over 2}M_3M_6\right)\ee
which transforms the VBS masses into the Neel masses. Note that the above unitary transformation commutes with $\G_i$ and the charge SU(2) generators. 

\section{The Berry phase of the space-time Neel hedgehog in the domain wall of the O(6) NL$\s$ model}\label{elaboration}

The statement that the space-time Neel hedgehog in the domain wall of the O(6) NL$\s$ model has Berry's phase ${1\over i} w^\dagger\p_z w$ requires a bit more explanation. This is because the hedgehog configuration exists only within thickness $d$ around the domain wall. To argue that we can use the result in ``The Neel hedgehog'' section of the main text, consider the following. (1) Consider the hedgehog configuration where the VBS order parameter is $v_3={\rm sign}(z)$ in the core for $|z|>d$. For such configuration, the Berry phase is zero outside the domain wall region. This is because for $|z|>d$ only four out of the six components of $\tilde{\hat{\Omega}}$ is non-zero, and by the same argument used in determining the statistics of the VBS hedgehog in the main text we conclude the Berry phase is zero. (2) For $|z|>d$ we can smoothly deform the hedgehog order parameter configuration so that $\Omega=(0,0,0,0,0,{\rm sign}(z))$.

\bibliographystyle{ieeetr}
\bibliography{bibs_O6}